\documentclass[english,aps, superscriptaddress, floatfix, 12pt]{revtex4}
\usepackage[T1]{fontenc}
\usepackage[latin1]{inputenc}
\usepackage{graphicx}
\usepackage{amssymb}
\usepackage{epsfig}

\makeatletter

\usepackage{babel}
\makeatother
\begin{document}

\title{Following Gluonic World Lines to Find\\ 
the QCD Coupling in the Infrared}

\author{Dmitri Antonov, Hans-J\"urgen Pirner\\
{\it Institut f\"ur Theoretische Physik, Universit\"at Heidelberg,\\
Philosophenweg 19, D-69120 Heidelberg, Germany}}

\begin{abstract}
Using a parametrization of the Wilson loop with the minimal-area law,
we calculate the polarization operator of a valence gluon, which propagates in the confining background. This enables us to obtain
the infrared freezing (i.e. finiteness) of the running strong coupling in the confinement phase, as well as in the  
deconfinement phase up to the temperature of dimensional reduction. The momentum scale defining the onset of freezing 
is found both analytically and numerically. 
The nonperturbative contribution to the thrust variable, originating from the freezing, makes the value of this variable closer
to the experimental one. 
\end{abstract}

\maketitle

\section{Introduction}
The path-integral representation for the Green function of a particle moving along a closed trajectory and interacting with the gauge field yields the Wilson loop. When the particle is confined by the gauge field, its Wilson loop obeys the area law. 
In this physically very important case the path integral cannot be calculated analytically, because finding the minimal surface for an arbitrary contour $x_\mu(\tau)$ in $d>2$ dimensions is too complicated. Therefore, effective parametrizations of the minimal surface have been invented in the literature (see e.g.~\cite{param, t}), which means there exist certain physically motivated tricks
to construct the minimal-area functional in terms of $x_\mu(\tau)$. 

To give an example we consider the following formula, which converts a double surface integral into line integrals
with Stokes' theorem:
\begin{equation}
\label{e}
\int_\Sigma^{} d\sigma_{\mu\nu}(x)\int_\Sigma^{} d\sigma_{\mu\rho}(x')
\partial_\nu\partial_\rho D(x-x')=
-\oint_C^{} dx_\mu\oint_C^{} dx'_\mu D(x-x').
\end{equation}
Here $D$ is an arbitrary function for which the integrals are finite, $\partial_\nu\equiv\frac{\partial}{\partial x_\nu}$, and $\Sigma$ is an arbitrary surface encircled by the contour $C$. As one can see, the choice $D(x-x')=(x-x')^2$ is the unique one, for which the derivatives on the L.H.S. of Eq.~(\ref{e}) are removed completely, and one obtains 
$$\int_\Sigma^{} d\sigma_{\mu\nu}(x)\int_\Sigma^{} d\sigma_{\mu\nu}(x')=\oint_C^{} dx_\mu\oint_C^{} dx'_\mu x_\nu x'_\nu.$$ 
Now, if $C$ is a flat contour, then we can choose $\Sigma$ flat as well, in which case 
the L.H.S. of this equation equals to $2S^2$, where $S$ is the area of $\Sigma$. Therefore, for a flat contour, the minimal area reads 
\begin{equation}
\label{example}
S=\sqrt{\frac12\Sigma_{\mu\nu}^2},~~ {\rm where}~~  
\Sigma_{\mu\nu}\equiv\oint_C^{}dx_\mu x_\nu
\end{equation}
is the so-called tensor area~\cite{t}, which is manifestly a functional of $C$ only. Flat contours are a good approximation for particle trajectories when the particle is heavy. For example, the  
parametrization of Eq.~(\ref{example}) reproduces correctly the heavy-quark condensate~\cite{c} and the mixed heavy-quark--gluon condensate~\cite{c1} in QCD.
However, it certainly cannot be correct for a light particle, whose trajectory may deviate significantly from the flat one.

A parametrization for light and even massless 
particles has been proposed in Ref.~\cite{yuas}. Similarly to Eq.~(\ref{example}), it allows to express the area $S_{\rm min}$ of the minimal surface $\Sigma$ in terms of a single integral. 
The main idea is to convert the proper time in the path integral to a length coordinate $\tau\in[0,R]$. After that, one can naturally parametrize $S_{\rm min}$ as an integral of the transverse direction, $|{\bf r}(\tau)|$, along this coordinate:
\begin{equation}
\label{sMin}
S_{\rm min}=\int_0^R d\tau|{\bf r(\tau)}|.
\end{equation} 

With such a parametrization of $S_{\rm min}$, we calculate in this paper the  
polarization operator of a gluon when it splits into two valence gluons. The latter move in a nonperturbative background, which 
confines them. In the absence of the confining background, the polarization operator yields the standard Yang-Mills one-loop running coupling~\cite{book}.
In the presence of the background, the running coupling goes to a constant in the infrared region, i.e. its logarithmic growth "freezes"~\cite{yuas}: 
\begin{equation}
\label{fr}
\alpha_s(p)=\frac{4\pi}{b\ln\frac{p^2}{\Lambda^2}}\to
\frac{4\pi}{\tilde b\ln\frac{p^2+m^2}{\Lambda^2}}.
\end{equation}
Here, 
$$b=\frac{11}{3}N_c$$ 
is the absolute value of the first coefficient of the Yang-Mills $\beta$-function, 
$m\propto({\rm string~ tension})^{1/2}$ is a nonperturbative mass parameter.

In this paper, we will show how freezing occurs and calculate the values of $m$ and $\tilde b$ both analytically and numerically. 
The paper is organized as follows. In section~2, we recall the 
derivation of $\alpha_s(p^2)$ in the absence of the confining background, introducing the scalar polarization 
operator $\Pi_{\rm free}(p^2)$. In section~3, the world-line integrals for 
the polarization operator $\Pi(x,y)$ in the coordinate representation are evaluated. In section~4, freezing is discussed in detail.
In section~5, we extend this approach to the analysis of freezing in the deconfined phase. In section~6, possible
phenomenological consequences of freezing are discussed. Finally, the main results of the paper are summarized in Conclusions.

\section{Polyakov's derivation of the running strong coupling}
In this section, we recollect some steps of the derivation of $\alpha_s(p^2)$ based on the integration over 
quantum fluctuations of the Yang-Mills field~\cite{book} (see also~\cite{ps}). The procedure starts with splitting 
the total Yang-Mills field $A_\mu^a$ into a background, $\bar A_\mu^a$, and a quantum fluctuation, $a_\mu^a$, whose momentum is 
larger than that of the background. One can therefore substitute the Ansatz $A_\mu^a=\bar A_\mu^a+a_\mu^a$ into the bare Yang-Mills action, 
$$S_0[A]=\frac{1}{4g_0^2}\int d^4x(F_{\mu\nu}^a[A])^2,$$ 
where $F_{\mu\nu}^a[A]$ is the QCD field-strength tensor and $g_0$ the bare coupling.
One separates $A_\mu^a$ into a slowly varying background field $\bar A_\mu^a$ and fluctuations $a_\mu^a$:
$$A_\mu^a=\bar A_\mu^a+a_\mu^a.$$
Fixing the so-called background Feynman gauge $(D_\mu a_\mu)^a=0$, one adds to the action the term
$S_{\rm g.f.}=\frac{1}{2g_0^2}\int d^4x\left[(D_\mu a_\mu)^a\right]^2$, where 
$(D_\mu a_\nu)^a=\partial_\mu a_\nu^a+f^{abc}\bar A_\mu^b a_\nu^c$. The total gluon action reads
$$S_0+S_{\rm g.f.}=$$
$$=\frac{1}{4g_0^2}\int d^4x\left\{(F_{\mu\nu}^a[\bar A])^2-4a_\nu^a(D_\mu F_{\mu\nu}[\bar A])^a-2a_\mu^a\left[\delta_{\mu\nu}
(D^2)^{ac}+2f^{abc}F_{\mu\nu}^b[\bar A]\right]a_\nu^c+{\cal O}(a^3)\right\}.$$
To perform the one-loop renormalization of $g^2$, one has to integrate out the 
$a_\mu^a$-gluons in the $\bar A_\mu^a$-background. 
The full renormalized effective action can be written as 
\begin{equation}
\label{fullaction}
S=S_0+S^{\rm dia}+S^{\rm para}=\int\frac{d^4p}{(2\pi)^4}\frac{1}{4g^2(p)}F_{\mu\nu}^a(p)F_{\mu\nu}^a(-p),
\end{equation}
where the running coupling $g(p)$ is the object of the calculation.
The so-called diamagnetic part of the  
effective action for $\bar A_\mu^a$-fields has the form
\begin{equation}
\label{tr}
S^{\rm dia}={\rm tr}{\,}\ln(-D^2)={\rm tr}{\,}\left[(-\partial^2)^{-1}_{xx}\Delta^{(2)}(x)-
\frac12(-\partial^2)^{-1}_{xy}\Delta^{(1)}(y)(-\partial^2)^{-1}_{yx}\Delta^{(1)}(x)\right],
\end{equation}
where we have used the decomposition 
$$
-D^2=-\partial^2+\Delta^{(1)}+\Delta^{(2)}~~ {\rm with}~~ 
\Delta^{(1)}(x)\equiv it^a\left(\partial_\mu\bar A_\mu^a+2\bar A_\mu^a\partial_\mu\right),~~  
\Delta^{(2)}(x)\equiv (\bar A_\mu^at^a)^2,
$$
$(t^a)^{bc}=-if^{abc}$. The paramagnetic part of the effective action reads
\begin{equation}
\label{tr1}
S^{\rm para}=\frac12{\,}{\rm tr}{\,}\ln\left[1+(-\partial^2)^{-1}(2f^{abc}F_{\mu\nu}^b)\right]=
-\frac14(-\partial^2)^{-1}_{xy}(2f^{abc}F_{\mu\nu}^b(y))(-\partial^2)^{-1}_{yx}(2f^{adc}F_{\mu\nu}^d(x)).
\end{equation}
The subscript "dia" describes the diamagnetic interaction 
of the $\bar A_\mu^a$-field with the orbital motion of the $a_\mu^a$-gluons. This effect leads to the screening of charge and 
is present in the Abelian case as well. The subscript "para" is because this part of the effective action describes the paramagnetic interaction of the $\bar A_\mu^a$-field with the spin of of the $a_\mu^a$-gluons. It leads to the antiscreening of charge, which is a specific property of the non-Abelian gauge theories.
The dia- and paramagnetic parts of the one-loop effective action can be written as
\begin{equation}
\label{o}
S^{\left\{{\rm dia}\atop {\rm para}\right\}}=N_c\cdot\left\{
\frac{1}{12}\atop (-1)\right\}\cdot\int\frac{d^4p}{(2\pi)^4}F_{\mu\nu}^a(p)F_{\mu\nu}^a(-p)\Pi_{\rm free}(p^2)
\end{equation}
where
\begin{equation}
\label{pinofr}
\Pi_{\rm free}(p^2)\equiv\frac{1}{16\pi^2}\ln\frac{\Lambda_0^2}{p^2}=\int\frac{d^4q}{(2\pi)^4}\frac{1}{q^2(q+p)^2}
\end{equation}
is the free scalar polarization operator. Equation~(\ref{fullaction}) then yields  
\begin{equation}
\label{Renorm}
\frac{1}{g^2(p)}=\frac{1}{g_0^2}-\frac{b}{16\pi^2}\ln\frac{\Lambda_0^2}{p^2},
\end{equation}
where $g_0\equiv g(\Lambda_0)$.
Introducing the renormalized cutoff $\Lambda=\Lambda_0\exp\left(-\frac{8\pi^2}{bg_0^2}\right)$,
one finally arrives at the standard result $\alpha_s(p)=\frac{4\pi}{b\ln\frac{p^2}{\Lambda^2}}$.

\section{Calculation of the polarization operator}

In reality, the gluon fluctuations $a_\mu^a$'s do not appear as measurable excitations in the QCD spectrum and therefore must be self-confined, which means that the two $a_\mu^a$-gluons propagating along the loop interact and form a colored bound state. This interaction 
has a one-gluon-exchange part plus a nonperturbative part, which may be related to string formation in the octet-octet 
channel coupled to a color octet. If one assumes Casimir scaling for the string tension, the
corresponding string tension in the octet channel is related to that of the fundamental 
representation as (see e.g.~\cite{sz}) $\sigma=\frac98\sigma_{\rm fund}=0.225{\,}{\rm GeV}^2$ with 
$\sigma_{\rm fund}=0.2{\,}{\rm GeV}^2$.
The $a_\mu^a$-gluons may be confined because of a stochastic background field $B_\mu^a$, whose momenta are even smaller than the momenta of the $\bar A_\mu^a$-gluons~\cite{yuas} (see Fig.~1). The presence of this additional 
background field can be taken into account by substituting into the Yang-Mills action a modified Ansatz 
$$A_\mu^a=B_\mu^a+\bar A_\mu^a+a_\mu^a.$$ 
Accordingly, the definition of the one-loop effective action includes now the 
average over the background: $S=-\ln\left<\int{\cal D}a_\mu^a{\rm e}^{-S[A]}\right>_B$, where 
$\left<\ldots\right>_B$ is some gauge- and $O(4)$-invariant integration measure. In the course of the calculation of 
$\int{\cal D}a_\mu^a{\rm e}^{-S[A]}$, the presence of the background leads to the substitution $\partial^2\to D^2[B]$ in Eqs.~(\ref{tr}) and~(\ref{tr1}), where $(D_\mu[B] a_\nu)^a=\partial_\mu a_\nu^a+f^{abc}B_\mu^b a_\nu^c$. 
In the diagrammatic language, we are still considering the same one-loop diagrams, which contain {\it only two} 
external lines of the $\bar A_\mu^a$-field, {\it but} with the internal $a_\mu^a$-loop receiving {\it infinitely many} contributions of the $B_\mu^a$-field. The latter appear through the path-integral representation of the 
operator $(D^2[B])^{-1}$, which contains the Wilson line of the $a_\mu^a$-gluon in the $B_\mu^a$-field.
Such an $a_\mu^a$-gluon is called valence gluon from now on.

\begin{figure}
\includegraphics[clip,scale=0.4]{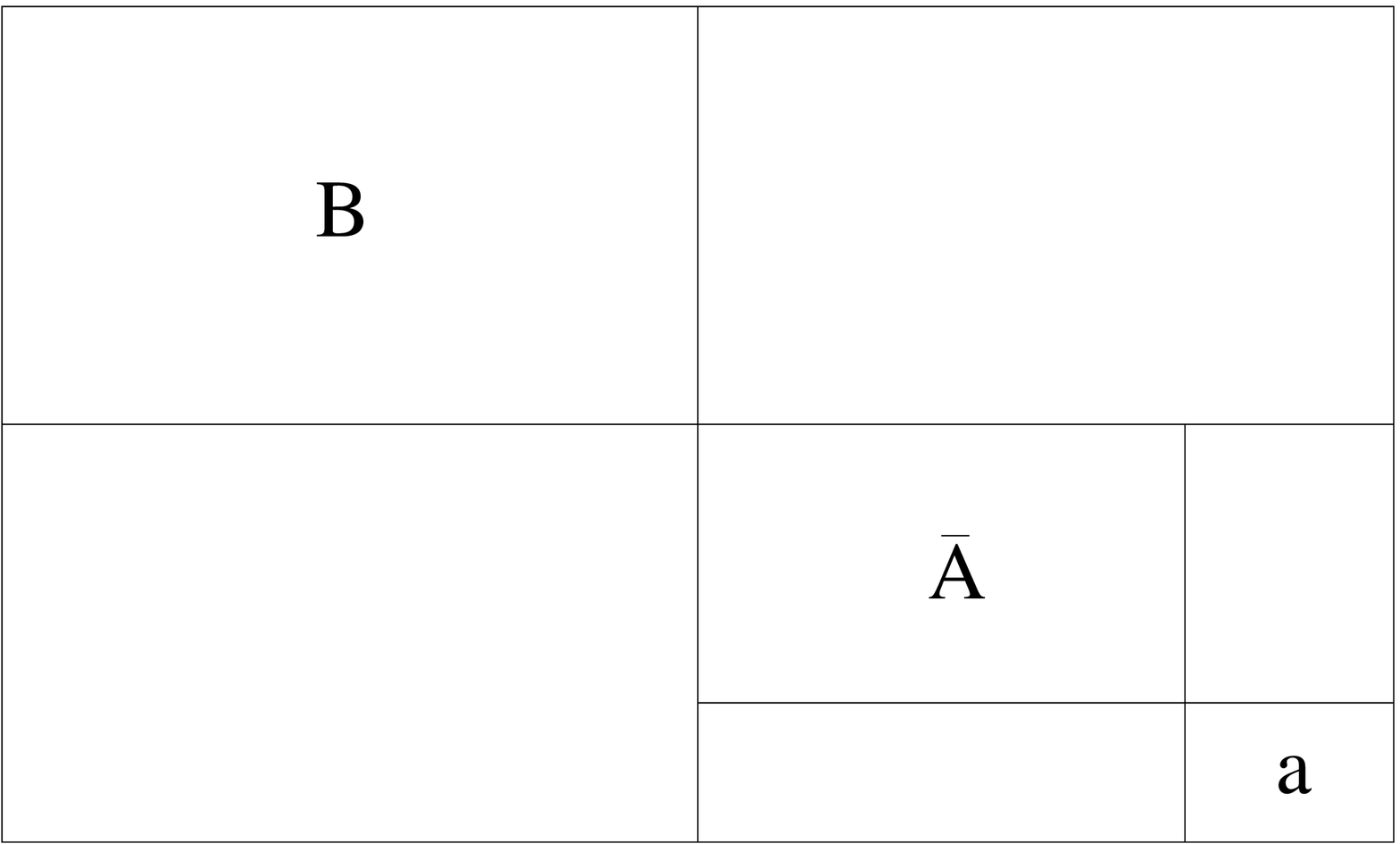}
\caption{Relative wavelengths of the three fields $B_\mu^a$, $\bar A_\mu^a$, and $a_\mu^a$.}
\end{figure}

In the coordinate representation the scalar polarization operator has the general form  
\begin{equation}
\label{frsc}
\Pi_{\rm free}(x)=\int\frac{d^4p}{(2\pi)^4}{\rm e}^{ipx}\Pi_{\rm free}(p^2)=D_0^2(x),~~ {\rm where}~ 
D_0(x)=\frac{1}{4\pi^2x^2}.
\end{equation}
In an arbitrary background $B_\mu^a$ the polarization operator becomes
$$\left<\Pi(x,y|B)\right>_B=\left<{\rm tr}{\,}(D^2[B])^{-1}_{xy}(D^2[B])^{-1}_{yx}\right>_B.$$ 
The path-integral representation of this average reads (see e.g.~\cite{yuas})
$$\left<\Pi(x,y|B)\right>_B\equiv\Pi(x,y)=\int_0^\infty ds\int_0^\infty d\bar s\int({\cal D}z_\mu)_{xy}({\cal D}\bar z_\mu)_{yx}
\exp\left(-\int_0^s d\lambda\frac{\dot z_\mu^2}{4}-\int_0^{\bar s}d\bar\lambda\frac{\dot{\bar z}_\mu^2}{4}\right)\times$$
\begin{equation}
\label{eQ}
\times\left<{\rm tr}{\,}{\cal P}{\,}\exp\left[i\left(\int_0^s d\lambda\dot z_\mu B_\mu^a(z)t^a+\int_0^{\bar s} d\lambda\dot{\bar z}_\mu B_\mu^a(\bar z)t^a\right)\right]\right>_B.
\end{equation}
Here, $({\cal D}z_\mu)_{xy}$ is the standard measure of integration over all paths $z_\mu(\lambda)$ such that $z(0)=y$, $z(s)=x$.
Namely, in $d$ dimensions,
$$({\cal D}z_\mu)_{xy}=\lim\limits_{N\to\infty}^{}\prod\limits_{k=1}^{N}\int\frac{d^dz_k}{(4\pi\varepsilon)^{d/2}},~~ 
z_k\equiv z(k\varepsilon),~~ \varepsilon=s/N.$$
The $B_\mu^a$-averaged term in Eq.~(\ref{eQ}) is the Wilson loop of the valence gluon. The confinement of the latter is reflected in the so-called area law. The Wilson loop $\left<\ldots\right>_B$ can be approximated by
${\rm e}^{-\sigma S_{\rm min}}$, where $S_{\rm min}=S_{\rm min}[z,\bar z]$ is the area of the minimal surface encircled by the paths $z_\mu(s)$ and $\bar z_\mu(\bar s)$. Moreover, parametrization of $S_{\rm min}$ in the form of~(\ref{sMin}) leads finally to the restoration of the translation invariance of $\Pi(x,y)$.

To calculate $\Pi(x,y)$ we introduce "center-of-mass" and relative coordinate of the gluons~\cite{yuas}: 
$$u_\mu=\frac{\bar sz_\mu+s\bar z_\mu}{s+\bar s},~~ r_\mu=z_\mu-\bar z_\mu.$$
We define new integration variables, which have the dimension of mass:
\begin{equation}
\label{mubarmu}
\mu=\frac{|x-y|}{2s},~~ \bar\mu=\frac{|x-y|}{2\bar s}.
\end{equation}
Then the kinetic terms of the gluons become
$$\int_0^s\frac{\dot z_\mu^2}{4}d\lambda+\int_0^{\bar s}\frac{\dot{\bar z}_\mu^2}{4}d\bar\lambda=
\frac12\int_{0}^{R}d\tau\left[(\mu+\bar\mu)\dot u_\mu^2+\mu_r\dot r_\mu^2\right]$$ 
with
$$R\equiv|x-y|.$$
Here, $\mu_r\equiv\frac{\mu\bar\mu}{\mu+\bar\mu}$ is the "reduced mass", and $\tau$ is the distance to the point $x$ along the line passing through $x$ and $y$. Note that, unlike Schwinger's proper time $s$, which has the dimension $({\rm length})^2$, the variable $\tau$ has the dimension $({\rm length})$.
The minimal area, $S_{\rm min}$, can be
effectively parametrized with the coordinate $\tau$ alone, after which the problem reduces to that of a
Schr\"odinger equation with the potential $\sigma|{\bf r}|$. The main new idea
of the present paper is to use a path-integral approach for the calculation of $\Pi(x,y)$. To this end, we will apply the Cauchy-Schwarz inequality to Eq.~(\ref{sMin}):
\begin{equation}
\label{smin}
S_{\rm min}=\int_0^R d\tau|{\bf r}(\tau)|\le\left(R\int_0^R d\tau{\bf r}^2\right)^{1/2}.
\end{equation}

Note that, in case of a (1+1)-dimensional 
classical-mechanics problem, this approximation works with a good accuracy. For two
particles of mass $m$ interacting through a linear potential, which
move from $(y,t)=(0,0)$ to the point $(0,R)$, one can write the equation of motion $m\ddot y_1=-m\ddot y_2=-\sigma$. Substituting the solution, 
$y_1=-y_2=\frac{\sigma}{2m}t(R-t)$ into
the exact formula $S_{\rm min}=\int_0^R dt(y_1-y_2)$, we get $S_{\rm min}=\frac{\sigma R^3}{6m}$. Using instead 
our approximation, we obtain 
$S_{\rm min}\le\left[R\int\limits_0^R dt(y_1-y_2)^2\right]^{1/2}=\frac{\sigma R^3}{\sqrt{30}m}$.
The relative error is therefore quite small, namely $\left(\frac{1}{\sqrt{30}}-\frac16\right):\frac16\simeq0.096$.

Encouraged by this observation, we return to the 4d case and eliminate the square root in the Cauchy-Schwarz inequality, Eq.~(\ref{smin}), by introducing an integration over an auxiliary parameter $\lambda$, which is sometimes called einbein. The expression for the polarization operator then becomes
$$
\Pi(x,y)\simeq\frac{R^2}{4}\int_0^\infty\frac{d\mu}{\mu^2}\int_0^\infty\frac{d\bar\mu}{\bar\mu^2}
\int({\cal D}u_\mu)_{xy}({\cal D}r_\mu)_{00}\times$$
\begin{equation}
\label{PI}
\times\int_0^\infty\frac{d\lambda}{\sqrt{\pi\lambda}}\exp\left[-\lambda-
\frac{\mu+\bar\mu}{2}\int_0^R d\tau\dot u_\mu^2-\frac{\mu_r}{2}\int_0^R d\tau\dot r_\mu^2-
\frac{\sigma^2R}{4\lambda}\int_0^R d\tau{\bf r}^2\right].
\end{equation}
Now, the integrals over $u_\mu(\tau)$ and $r_4(\tau)$ 
are free path integrals, while the integral over ${\bf r}(\tau)$ is that of a harmonic oscillator. We calculate these integrals 
by introducing the variables 
$$\xi\equiv\sigma R^{3/2}/\sqrt{2\mu_r\lambda}~
{\rm instead~ of}~ \lambda,~~ a=\mu R/2~ {\rm instead~ of}~ \mu,~~ {\rm and}~ b=\bar\mu R/2~ {\rm instead~ of}~ \bar\mu.$$
In terms of these variables
$$\Pi(x,y)\equiv\Pi(R)=\frac{\sigma}{16\pi^{9/2}R^2}f(\sigma R^2),$$ 
where 
\begin{equation}
\label{eqf}
f(\sigma R^2)=\int_0^\infty\frac{d\xi}{\sqrt{\xi}\sinh^{3/2}\xi}
\int_0^\infty dadb\sqrt{\frac{a+b}{ab}}
\exp\left[-a-b-\left(\frac{\sigma R^2}{2\xi}\right)^2\frac{a+b}{ab}\right].
\end{equation}

\begin{figure}
\epsfig{file=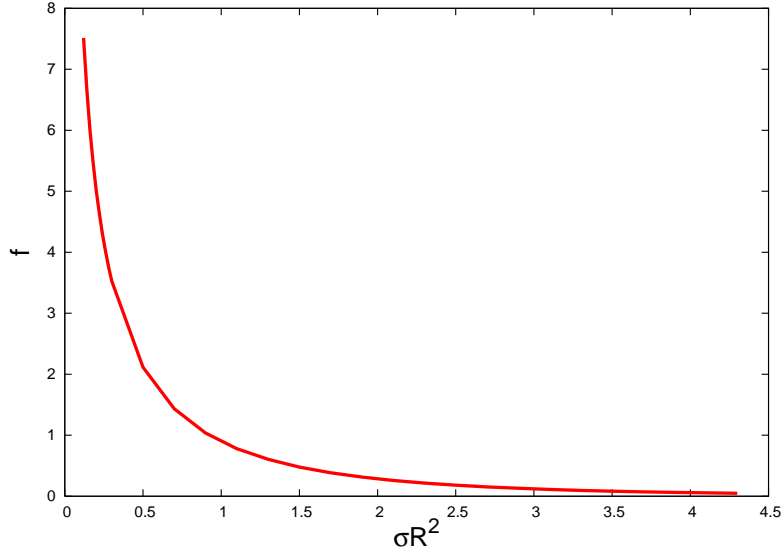, width=105mm}
\caption{Integral $f(\sigma R^2)$ from Eq.~(\ref{eqf}).}
\end{figure}

When integrating analytically over $a$ and $b$, it has been found (cf. Appendix~A) 
that, owing to the $a\leftrightarrow b$ symmetry, 
the corresponding saddle-point equations can be solved even when the pre-exponent is lifted to the exponent. However, the leading {\it large-distance} asymptotic behavior of $\Pi(R)$ stems from the mere substitution of the saddle-point values without that lifting, $a=b\simeq\frac{\sigma R^2}{2\xi}$, into the pre-exponent (see Appendix~A for details). This procedure yields
\begin{equation}
\label{Papr}
\Pi(R)\simeq\frac{\sigma^{3/2}}{16\pi^{7/2}R}\int_0^\infty\frac{d\xi}{\xi\sinh^{3/2}\xi}{\rm e}^{-2\sigma R^2/\xi}.
\end{equation}
This integral can be evaluated by splitting the integration region into two parts: 
\begin{equation}
\label{split}
\Pi(R)\equiv\frac{\sigma^{3/2}}{16\pi^{7/2}R}(I_1+I_2),
\end{equation}
where 
$$I_1\simeq\int_0^1\frac{d\xi}{\xi^{5/2}}{\rm e}^{-2\sigma R^2/\xi},~~ I_2\simeq2^{3/2}\int_1^\infty\frac{d\xi}{\xi}
{\rm e}^{-\frac{3\xi}{2}-\frac{2\sigma R^2}{\xi}}.$$
Then, at $\sigma R^2\gg 1$, $I_1={\cal O}({\rm e}^{-2\sigma R^2})$ is a subleading term. The integral $I_2$ is saturated by its saddle point, $\xi=2R\sqrt{\sigma/3}$, which lies inside the integration region. This yields  
\begin{equation}
\label{large}
\Pi(R)\simeq\frac{\sigma^{5/4}}{2^{5/2}\cdot3^{1/4}\cdot\pi^3\cdot R^{3/2}}{\rm e}^{-2\sqrt{3\sigma}R}~~ {\rm at}~ 
\sigma R^2\gg 1.
\end{equation}

Let us now consider the opposite limit of small distances. There, Eq.~(A.3) goes over to 
$\left(\frac{\mu_r}{2\pi R}\right)^{3/2}$, and the product of the three path integrals, Eqs.~(A.1)-(A.3), yields
$$\int({\cal D}u_\mu)_{xy}({\cal D}r_\mu)_{00}\exp\left[
-\frac{\mu+\bar\mu}{2}\int_0^R d\tau\dot u_\mu^2-\frac{\mu_r}{2}\int_0^R d\tau\dot r_\mu^2-
\frac{\sigma^2R}{4\lambda}\int_0^R d\tau{\bf r}^2\right]\to$$
$$\to\left[\frac{\mu\bar\mu}{(2\pi R)^2}\right]^2\exp\left[-\frac{(\mu+\bar\mu)R}{2}\right]~~ 
{\rm at}~ \sigma R^2\ll 1.$$
Since this expression does not depend on $\lambda$ anymore, the remaining $\lambda$-integration in Eq.~(\ref{PI})
results in a factor of 1. Equation~(\ref{PI}) then goes over to the free scalar polarization operator, Eq.~(\ref{frsc}):
\begin{equation}
\label{sm}
\Pi(x,y)\to\frac{1}{64\pi^4R^4}\int_0^\infty d\mu\int_0^\infty d\bar\mu\exp\left[-\frac{(\mu+\bar\mu)R}{2}\right]=
\Pi_{\rm free}(x-y)~~ {\rm at}~ \sigma R^2\ll 1.
\end{equation}
Therefore, we have a formula for the polarization operator, interpolating between the limits~(\ref{large}) and~(\ref{sm}): 
\begin{equation}
\label{appr}
\Pi(R)=\frac{1}{16\pi^4R^4}{\rm e}^{-A\sqrt{\sigma R^2}}\left(1+B(\sigma R^2)^{5/4}\right),
\end{equation}
with the following analytic values of the coefficients: $A=2\sqrt{3}\simeq3.46$, 
$B=\frac{2^{3/2}\pi}{3^{1/4}}\simeq6.75$.

\section{Freezing of the running strong coupling at zero temperature}

The integral of Eq.~(\ref{eqf}) has been also calculated with the Monte-Carlo integration routine Vegas~\cite{vegas}
in the interval $0.12\le\sigma R^2\le4.3$.
The interpolating curve is plotted in Fig.~2. The numerical fit to these data yields $A=3.38$, which is very 
close to 3.46. The corresponding analytic and numerical values of the mass $m$ in Eq.~(\ref{fr}) are
\begin{equation}
\label{masses}
m_{\rm an}=2\sqrt{3\sigma}=1.64{\,}{\rm GeV},~~ m_{\rm num}=3.38\sqrt{\sigma}=1.60{\,}{\rm GeV}.
\end{equation}
Note that the value of the analytically calculated freezing mass depends on the dimensionality of space-time $d$ as
\begin{equation}
\label{ma}
m_{\rm an}=2\sqrt{(d-1)\sigma}.
\end{equation}
This is readily seen from the saddle-point of the integral $I_2$ in Eq.(\ref{split}) by noticing that, in $d$ dimensions, $\sinh^{3/2}\xi\to\sinh^{\frac{d-1}{2}}\xi$ in Eq.~(\ref{Papr}).
This result is a direct consequence of the Ansatz for the minimal area we use, Eq.~(\ref{smin}).

Freezing, as defined by Eq.~(\ref{fr}), stems from the replacement of the free-gluon polarization operator, 
Eq.~(\ref{pinofr}), by that of the valence gluon,
\begin{equation}
\label{piv}
\Pi_{\rm val}(p^2)\equiv\frac{1}{16\pi^2}\ln\frac{\Lambda_0^2}{p^2+m^2}.
\end{equation}
Therefore, it looks instructive to compare the inverse Fourier image of this desired exact expression with 
our approximate result, Eq.~(\ref{appr}). We have 
$$\int d^4p{\rm e}^{-ipx}\ln\frac{\Lambda_0^2}{p^2+m^2}=\int_0^\infty\frac{ds}{s}\int d^4p{\rm e}^{-ipx}
\left[{\rm e}^{-s(p^2+m^2)}-{\rm e}^{-s\Lambda_0^2}\right]=$$
$$=\pi^2\int_0^\infty\frac{ds}{s^3}{\rm e}^{-\frac{x^2}{4s}-m^2s}=
\frac{8\pi^2m^2}{x^2}K_2(m|x|),$$
where $K_2$ is a Macdonald function.
(When deriving the third formula in this chain, we have used the obvious fact that 
$\int d^4p{\rm e}^{-ipx}=0$ for $x\ne0$.) Therefore, Eq.~(\ref{piv}) in the coordinate representation reads
\begin{equation}
\label{im}
\Pi_{\rm val}(R)=\frac{m^2}{32\pi^4R^2}K_2(mR).
\end{equation}
The short-distance asymptotic limit of this formula coincides with Eq.~(\ref{sm}).
As for the large-distance limit, we see that Eq.~(\ref{im}) has the same exponential fall-off as our result, 
Eq.~(\ref{appr}), but a different pre-exponential $R$-behavior. The ratio of Eq.~(\ref{appr}) 
to Eq.~(\ref{im}) at $mR\gtrsim1$ is $\propto\sqrt{\sigma}R$. However, at large distances in question, this 
discrepancy is unimportant.

\begin{figure}
\epsfig{file=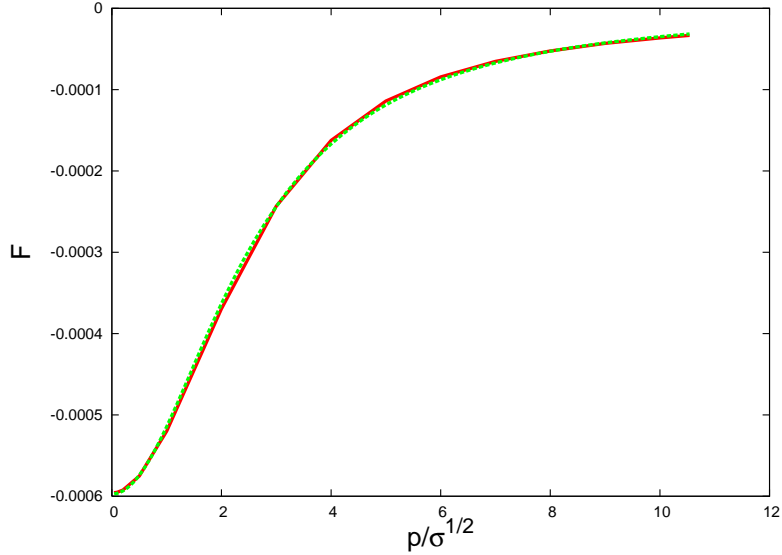, width=105mm}
\caption{The function $F(q)$ (red online) and a fit to it (green online).}
\end{figure}

Finally, it is worthwhile to compare directly Eq.~(\ref{piv}) with the Fourier image of Eq.~(\ref{appr}). Because of the 
logarithmic divergency, we compare the derivatives of the two expressions, which are UV-finite. Therefore, we 
compare $\frac{d\Pi_{\rm val}(p^2)}{dp^2}=-\frac{1}{16\pi^2}\frac{1}{p^2+m^2}$ with $\frac{d\Pi(p^2)}{dp^2}$, 
where $\Pi(p^2)$ is the Fourier image of $\Pi(R)$: $\Pi(p^2)=\frac{4\pi^2}{|p|}\int_0^\infty dRR^2J_1(|p|R)\Pi(R)$, 
$J_\nu$'s are the Bessel functions. Introducing the dimensionless variables $x=\sqrt{\sigma}R$ and $q=|p|/\sqrt{\sigma}$, one has
$$\frac{d\Pi(p^2)}{dp^2}\equiv\frac{F(q)}{\sigma},~~ {\rm where}$$ 
\begin{equation}
\label{func}
F(q)=\frac{1}{8\pi^2q^2}\int_0^\infty\frac{dx}{x}\left[\frac12\left(J_0(qx)-J_2(qx)\right)-\frac{J_1(qx)}{qx}\right]
\left(1+Bx^{5/2}\right){\rm e}^{-Ax}.
\end{equation}
This function has been calculated numerically for 
$0.02{\,}{\rm GeV}\le |p|\le 5{\,}{\rm GeV}$, which corresponds to $q\in[0.042, 10.541]$. Fitting the result by the function
\begin{equation}
\label{funcfit}
F_{\rm fit}(q)=-\frac{1}{16\pi^2}\frac{d_1}{q^2+d_2^2},
\end{equation} 
we obtain: $d_1=0.585$, $d_2=2.491$.
The interpolating curve for $F(q)$, at the above-mentioned values of $q$, and the function~(\ref{funcfit})
are plotted in Fig.~3. The value of $m$ corresponding to the coefficient $d_2$ is therefore
\begin{equation}
\label{mass1}
m=d_2\sqrt{\sigma}=1.18{\,}{\rm GeV}.
\end{equation}
It is closer to the phenomenological estimates ($\simeq 1{\,}{\rm GeV}$)~\cite{ph} than the values~(\ref{masses}). 
Furthermore, the coefficient $d_1$ defines a numerical prediction for the parameter $\tilde b$:
\begin{equation}
\label{momrep}
\tilde b\simeq 0.585 b=6.435.
\end{equation}
Therefore, the numerical analysis in the momentum representation yields an effective decrease of $b$ in the infrared region.
For the final form of $\alpha_s(p)$ at zero temperature, we refer the reader to the end of section~5, where we discuss this result 
together with the one at $T>T_c$.

\section{Freezing of the running strong coupling in the gluon plasma}
At temperatures above deconfinement, $T>T_c$, large spatial Wilson loops still exhibit the area law. 
For pure gauge SU(3) Yang-Mills theory, $T_c\simeq0.27{\,}{\rm GeV}$~\cite{Tc}. 
This behavior of spatial Wilson loops is the main well established nonperturbative phenomenon at $T>T_c$, which is 
usually called "magnetic" or "spatial" confinement~\cite{sp}. Of course, it does not contradict  
true deconfinement of a static quark-antiquark pair, since space-time Wilson loops indeed lose the exponential damping 
with the area for $T>T_c$. If points $x$ and $y$ 
are separated by a time-like interval, valence gluons in the polarization operator $\Pi(x,y|T)$ are not confined, and therefore $\alpha_s(p^2)$ at $p^2>0$ is given by the perturbative formula, without freezing. When $x$ and $y$
are separated by a space-like interval, magnetic confinement holds, and one expects freezing of $\alpha_s(p^2)$ at $p^2<0$.

We will start our analysis with the calculation of the spatial polarization operator 
of a valence gluon in the 
SU(3) pure Yang-Mills theory at temperatures {\it higher} than the temperature of dimensional reduction, 
$T>T_{\rm d.r.}\simeq2T_c$. QCD becomes a superrenormalizable theory in three spatial dimensions, where the renormalization of the dimensionful coupling, $g_3\equiv g\sqrt{T}$, is exact in one loop. Apparently, this effective coupling is not asymptotically free
and is unrelated to the study of freezing. The following calculation prepares the subsequent analysis of $\alpha_s$ at $T_c<T<T_{\rm d.r.}$.

Let us first consider the propagator of a free particle at temperature $T$ from the origin to the point $R_\mu=({\bf R},R_4)$:
$$
(-\partial^2)_{R,0}^{-1}=\int_0^\infty ds\sum\limits_{n}^{}\frac{1}{(4\pi s)^2}
\exp\left[-\frac{{\bf R}^2+(R_4-\beta n)^2}{4s}\right]=$$
\begin{equation}
\label{prop}
=\int_0^\infty ds\sum\limits_{n}^{}\frac{1}{2T\sqrt{\pi s}}\exp\left[-\frac{(R_4-\beta n)^2}{4s}\right]\cdot
T\int({\cal D}{\bf z})_{{\bf R}{\bf 0}}\exp\left(-\int_0^s\frac{\dot{\bf z}^2}{4}d\lambda\right),
\end{equation}
where $\sum\limits_{n}^{}\equiv\sum\limits_{n=-\infty}^{+\infty}$. 
Upon the Poisson resummation, one has 
\begin{equation}
\label{Poi}
\frac{1}{2T\sqrt{\pi s}}\sum\limits_{n}^{}\exp\left[-\frac{(R_4-\beta n)^2}{4s}\right]=\sum\limits_{n}^{}\exp\left(
-\omega_n^2s+i\omega_nR_4\right),~~ \omega_n=2\pi nT.
\end{equation}
When $T\to\infty$, only the zeroth term on the R.H.S. of Eq.~(\ref{Poi}) survives, 
which means dimensional reduction. The sum goes to 1, and
\begin{equation}
\label{propTemp}
(-\partial^2)_{R,0}^{-1}\to T\int_0^\infty ds\int({\cal D}{\bf z})_{{\bf R}{\bf 0}}\exp\left(-\int_0^s\frac{\dot{\bf z}^2}{4}d\lambda\right)=\frac{T}{4\pi L},
\end{equation}
where $L\equiv|{\bf R}|$.
With the effect of magnetic confinement included, the 
polarization operator for $T>T_{\rm d.r.}$ reads [cf. Eqs.~(\ref{eQ}) and (\ref{smin})]:
$$\Pi({\bf x},{\bf y}|T)=T^2\int_0^\infty ds\int_0^\infty d\bar s~ I(s,\bar s),~~ {\rm where}$$ 
\begin{equation}
\label{PiT}
I(s,\bar s)\simeq
\int({\cal D}{\bf z})_{{\bf x}{\bf y}}({\cal D}\bar{\bf z})_{{\bf y}{\bf x}}
\exp\Biggl[
-\int_0^s\frac{\dot{\bf z}^2}{4}d\lambda-\int_0^{\bar s}\frac{\dot{\bar{\bf z}}^2}{4}d\bar\lambda-
\sigma_s\Bigl(L\int_0^Ld\tau\vec\rho{\,}^2\Bigr)^{1/2}\Biggr].
\end{equation}
Here, $\sigma_s$ is the spatial string tension, whose ratio to the zero-temperature string tension, 
$\sigma_0=0.225{\,}{\rm GeV}^2$, is plotted in Fig.~4.
The points ${\bf x}$ and ${\bf y}$ are spatially separated, 
${\bf y}-{\bf x}={\bf R}$, and $\vec\rho$ is a two-dimensional vector orthogonal to ${\bf R}$. 
Two-dimensional vectors 
are denoted by an arrow, differently from three-dimensional vectors, which are boldfaced. 
This polarization operator can be calculated in a way similar to the zero-temperature one.
Referring the reader for details to Appendix~B, we present the final result: the
polarization operator at $T>T_{\rm d.r.}$ reads
\begin{equation}
\label{polTfinal}
\Pi({\bf x},{\bf y}|T)\simeq\frac{\sqrt{\sigma_s}T^2}{4\pi^2L}{\rm e}^{-mL}~~ {\rm at}~ \sigma_sL^2\gg 1,
\end{equation}

\begin{equation}
\label{PhighT}
\Pi({\bf x},{\bf y}|T)\simeq\left(\frac{T}{4\pi L}\right)^2~~ {\rm at}~ \sigma_sL^2\ll 1.
\end{equation}
In Eq.~(\ref{polTfinal}),
\begin{equation}
\label{MT}
m\equiv2\sqrt{2\sigma_s},
\end{equation}
i.e. Eq.~(\ref{ma}) at $d=3$ is reproduced. Equation~(\ref{PhighT}) is nothing but the free scalar polarization operator, that is just the square of Eq.~(\ref{propTemp}).

\begin{figure}
\epsfig{file=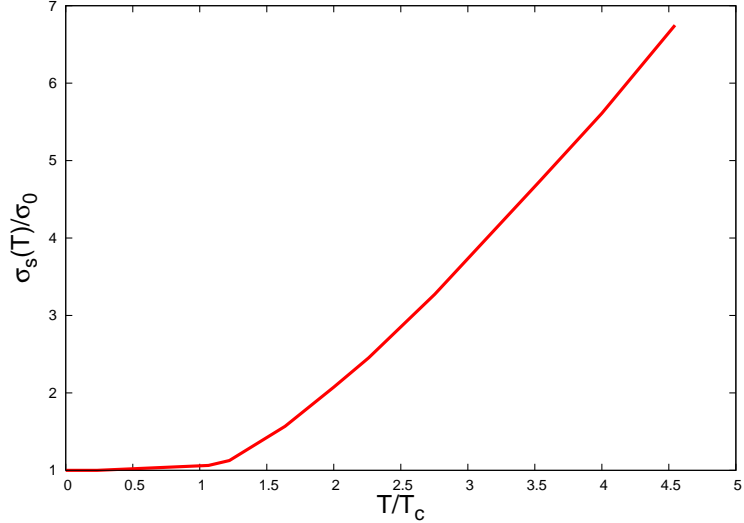, width=100mm}
\caption{The curve interpolating the lattice data on the ratio of the spatial string tension to the zero-temperature one in the SU(3) quenched QCD as a function of $T/T_c$~\cite{lat}.}
\end{figure}

Let us now proceed to the physically more interesting range of temperatures, $T_c<T<T_{\rm d.r.}$. Here, like at $T=0$, the 
exponential fall-off of $\Pi(x,y|T)$ (where $x$ and $y$ are four-vectors) due to magnetic confinement {\it is} relevant to the freezing of $\alpha_s$. The Euclidean path-integral representation for $\Pi(x,y|T)$ at $T_c<T<T_{\rm d.r.}$ can be constructed by
using Eqs.~(\ref{prop}) and (\ref{PiT}):
\begin{equation}
\label{genE}
\Pi(x,y|T)=\frac{1}{4\pi}\int_0^\infty\frac{ds}{\sqrt{s}}\int_0^\infty\frac{d\bar s}{\sqrt{\bar s}}
I(s,\bar s)\sum\limits_{n,k}^{}
\exp\left\{-\frac14\left[\frac{(R_4-\beta n)^2}{s}+\frac{(R_4-\beta k)^2}{\bar s}\right]\right\}.
\end{equation}
This quantity is calculated in Appendix~B. At large distances, $\sigma_sL^2\gg 1$, the result reads:
\begin{equation}
\label{ld}
\Pi(x,y|T)\simeq\frac{\sigma_sL}{8\sqrt{2}\pi^3R^3}{\rm e}^{-2\sqrt{2\sigma_sLR}},
\end{equation}
where $R\equiv\sqrt{R_4^2+L^2}$.
Recalling that, in the physical Minkowski space-time, magnetic confinement holds only when $R_\mu$ is a space-like vector, 
one can always place the points $x$ and $y$ along some spatial axis, which makes $L$ and $R$ equal. The resulting formula for the polarization operator at large distances takes the form
\begin{equation}
\label{Plar}
\Pi(x,y|T)\simeq\frac{\sigma_s}{8\sqrt{2}\pi^3R^2}{\rm e}^{-2\sqrt{2\sigma_s}R}~~ {\rm at}~ \sigma_sR^2\gg 1.
\end{equation}
Therefore, at the temperatures $T_c<T<T_{\rm d.r.}$ and Minkowskian $p^2<0$, freezing of $\alpha_s(p^2)$ takes place at the 
temperature-dependent momentum scale, which is analytically defined as $m=2\sqrt{2\sigma_s}$. The factor "2" under the square root in this formula is the number of spatial dimensions minus one, in accordance with Eq.~(\ref{ma}).

The limit of small distances is also discussed in Appendix~B. At $L=R$, the result reads
\begin{equation}
\label{pOl}
\Pi(x,y|T)\to\left[\frac{T}{4\pi R}\coth(\pi TR)\right]^2~~ {\rm at}~ \sigma_sR^2\ll 1.
\end{equation}
(In particular, at $T\to\infty$, this result goes over to Eq.~(\ref{PhighT}), as it should do.) 
Since $\sqrt{\sigma_0}=474{\,}{\rm MeV}$, $T_c=270{\,}{\rm MeV}$, one can see from Fig.~4  
that, at $T_c<T<T_{\rm d.r.}$, the condition $\sigma_sR^2\ll 1$ automatically means also 
$TR\ll 1$. For this reason, at these temperatures, Eq.~(\ref{pOl}) can be approximated by its zero-temperature counterpart,
$\frac{1}{(4\pi^2R^2)^2}$. 
Therefore, the formula for the polarization operator, which interpolates between this short-distance limit and the large-distance one, Eq.~(\ref{Plar}), reads
$$\Pi(x,y|T)=\frac{{\rm e}^{-2\sqrt{2\sigma_s}R}}{(4\pi^2R^2)^2}\left(1+\pi\sqrt{2}\sigma_sR^2\right)~~ 
{\rm at}~ T_c<T<T_{\rm d.r.}.$$
Note that this expression depends on temperature only implicitly, namely through $\sigma_s(T)$.

\begin{figure}
\epsfig{file=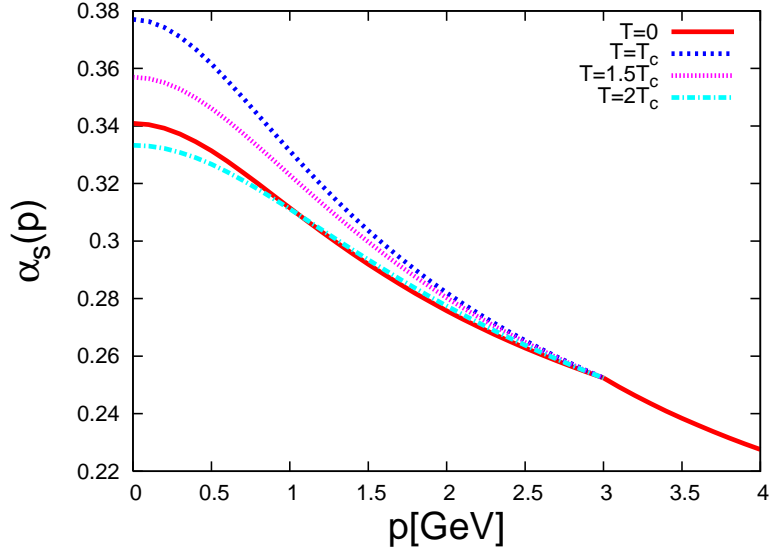, width=105mm}
\caption{The running coupling with freezing at $0{\,}\le p\le3{\,}{\rm GeV}$ for $T=0$, $T=T_c$, $T=1.5T_c$, and $T=2T_c$. 
The curve at $3{\,}{\rm GeV}\le p\le4{\,}{\rm GeV}$ is the experimental $\alpha_s(p)$, according to the web-site 
http://www-theory.lbl.gov/\~{\,}ianh/alpha/alpha.html from Ref.~\cite{pdg}.}
\end{figure}

The analysis of the polarization operator in the momentum representation can again be performed. Specifically, the factor 
$\left(1+Bx^{5/2}\right){\rm e}^{-Ax}$ in Eq.~(\ref{func}) should be replaced by 
$\left(1+\pi\sqrt{2}x^2\right){\rm e}^{-2\sqrt{2}x}$. Fitting the integral by the function~(\ref{funcfit}), we get the values
$d_1=0.711$, $d_2=2.289$. The freezing mass grows with temperature as the square root of $\sigma_s(T)$: $m(T)=d_2\sqrt{\sigma_s(T)}$. When $T$ varies from $T_c$ to $T_{\rm d.r.}\simeq2T_c$, $m(T)$ varies from 1.09~GeV to
1.56~GeV. Finally, the parameter $\tilde b\simeq 0.711b=7.821$ is larger than the one at zero-temperature, Eq.~(\ref{momrep}).

Both at $T=0$ and $T_c<T<T_{\rm d.r.}$, freezing modifies Eq.~(\ref{Renorm}) as 
$$\frac{1}{g^2(p)}=\frac{1}{g_0^2}-\frac{\tilde b}{16\pi^2}\ln\frac{\Lambda_0^2+m^2}{p^2+m^2}.$$
Defining the renormalized cutoff, $\Lambda$, through the bare one, $\Lambda_0$, as 
$$\Lambda=\sqrt{\Lambda_0^2+m^2}\exp\left(-\frac{2\pi}{\alpha_s(\Lambda_0)\tilde b}\right),$$ 
one enforces 
$\alpha_s(p)=\frac{4\pi}{\tilde b\ln\frac{p^2+m^2}{\Lambda^2}}$ to take the value $\alpha_s(\Lambda_0)$ when $p=\Lambda_0$.
It is then natural to choose a sufficiently large momentum scale $\Lambda_0$, 
where $\alpha_s$ is practically unaffected by
freezing and finite-temperature effects, and match it at that scale to the experimental value. Choosing $\Lambda_0=3{\,}{\rm GeV}$,
where $\alpha_s\simeq0.2524$, we plot in Fig.~5 $\alpha_s(p)$ with freezing at $T=0$, $T=T_c$, $T=1.5T_c$, and $T=2T_c$ for $p\le 3{\,}{\rm GeV}$, as well as the experimentally
measured $\alpha_s(p)$~\cite{pdg} for $3{\,}{\rm GeV}\le p\le4{\,}{\rm GeV}$. At at fixed $p$, one observes an increase of $\alpha_s(p)$ at $T=T_c$ with respect to $\alpha_s(p)$ at $T=0$, and a subsequent decrease with the growth of $T$.

\section{Possible phenomenological tests of the infrared freezing of the QCD coupling}
It has been pointed out in various papers ~\cite{Dokshitzer:1995qm, D} 
that the infrared behavior of the QCD running coupling can be used to
estimate power-behaved corrections to various QCD observables. The essential parameters
are few moments of the coupling in the infrared region.
One class of the possible observables consists of event-shape variables like the thrust ${\cal T}$:
$$
{\cal T}=\max\limits_{{\bf n}}\frac{\sum_i |{\bf p}_i {\bf n}|}{\sum_i |{\bf p}_i|}.$$
Here the vectors ${\bf p}_i$'s are the momenta of the final-state hadrons and ${\bf n}$ is an
arbitrary vector, which maximizes ${\cal T}$. If the momenta of the hadrons
form an almost collinear jet then, after the maximization, ${\bf n}$ will lie along the
jet axis. The value of the above thrust variable becomes 1 for an idealized pencil-like jet. 
Due to radiation of gluons, the observed ${\cal T}$ will be differing from 1. In 
perturbative QCD, these corrections from hard-gluon radiation at a certain scale $p$ can be calculated~\cite{ESW}. 
In addition to gluon radiation in the perturbative region, also
soft-gluon radiation is present below a scale $\mu_{\rm IR}\simeq 3{\,}{\rm GeV}$. Essentially the physics in this
region is supposed to be parametrized by the freezing of $\alpha_s$, which we derived in this
paper. Because of confinement, it seems to be impossible to map out this 
infrared running of the coupling exactly, i.e. for each momentum value. But measurements
about the energy loss in fragmentation may allow to access an integral over the
infrared region. The observable $1-{\cal T}$ related to thrust ${\cal T}$ has an expansion in terms of the perturbative 
$\alpha_s^{\rm pert}(p)$:
$$1-{\cal T}\Bigr|_{\rm pert}=0.334\alpha_s^{\rm pert}(p)+1.02 (\alpha_s^{\rm pert}(p))^2+
{\cal O}\left((\alpha_s^{\rm pert}(p))^3\right),$$
which acquires a hadronization correction due to soft gluon radiation
\begin{equation}
\label{FV}
1-{\cal T}= 1-{\cal T}\Bigr|_{\rm pert}+\frac {2 \lambda}{p}.
\end{equation}
The nonperturbative higher-twist contribution $\lambda$ may be 
related to an integral over the infrared region of $\alpha_s(p)=\frac{4\pi}{\tilde b\ln\frac{p^2+m^2}{\Lambda^2}}$ 
with freezing~\cite{D}, 
$$\lambda=C_F \int_0^{3{\,}{\rm GeV}} dp \frac{\alpha_s(p)}{\pi},$$
where $C_F=4/3$ is the Casimir operator of the fundamental representation of the group SU(3).
Using for $\alpha_s(p)$ the parameters $m$ and $\tilde b$ at $T=0$ and $T=T_c$, we obtain
\begin{equation}
\label{ourl}
\lambda\Bigr|_{T=0}=0.376{\,}{\rm GeV},~~ \lambda\Bigr|_{T=0.27{\,}{\rm GeV}}=0.395{\,}{\rm GeV}.
\end{equation}
In order to reproduce the experimental data on the energy losses through radiation in the fragmentation 
process of $c$- and $b$-quarks, a value $\lambda\Bigr|_{T=0}=0.5{\,}{\rm GeV}$ has been estimated~\cite{ESW}.

As a possible phenomenological application of freezing, let us 
evaluate the thrust of a two-jet event at the scale $p=M_Z$. Using the value 
$\alpha_s^{\rm pert}(M_Z)=0.12$~\cite{pdg}, one obtains the purely perturbative contribution
\begin{equation}
\label{thrpert}
1-{\cal T}\Bigr|_{\rm pert}=0.334\alpha_s^{\rm pert}(M_Z)+1.02 (\alpha_s^{\rm pert}(M_Z))^2\simeq0.055,
\end{equation}
which underestimates the experimental value measured at LEP~\cite{LEP} $1-{\cal T}\simeq0.068$.
Inclusion of the higher-twist contribution according to Eq.~(\ref{FV}), with the parameter $\lambda$ given by Eq.~(\ref{ourl}),  yields for the full quantity
\begin{equation}
\label{thr0}
1-{\cal T}\Bigr|_{T=0}=0.063,
\end{equation}
which is closer to the above-cited experimental value.

If hadronization takes place in the quark-gluon plasma, then one can study the effect of a modified
$\alpha_s(p)$ by investigating the energy loss due to soft radiation in the plasma. Accounting for the purely radiative higher-twist effect by means of 
$\lambda\Bigr|_{T=0.27{\,}{\rm GeV}}$ from Eq.~(\ref{ourl}), one gets at the $M_Z$-scale
\begin{equation}
\label{thrT}
1-{\cal T}\Bigr|_{T=0.27{\,}{\rm GeV}}=0.064.
\end{equation}
If one instead uses $\alpha_s(p)$ from Refs.~\cite{gbp}, then one gets
$\lambda\Bigr|_{T=0}=1.03{\,}{\rm GeV}$, $\lambda\Bigr|_{T=0.27{\,}{\rm GeV}}=0.61{\,}{\rm GeV}$, which overestimate 
the corresponding values~(\ref{ourl}), as well as the experimental value $\lambda\Bigr|_{T=0}=0.5{\,}{\rm GeV}$~\cite{ESW}.

In conclusion of this section, although $\alpha_s(p)$ with freezing enhances the perturbative contribution~(\ref{thrpert}) by 
15\%, the corresponding full values, Eqs.~(\ref{thr0}) and~(\ref{thrT}), are still smaller than the experimental one.
An increase of $\alpha_s(p)$ at $T=T_c$, with respect to the zero-temperature case, enhances the energy loss in the quark-gluon plasma, which occurs due to the infrared radiation. However, the amount of this enhancement, 
quantified by $(1-{\cal T})$, is only 1.6\%.

\section{Conclusions}
This paper analyses the so-called IR freezing (i.e. finiteness) of the running strong coupling, in the 
confinement and deconfinement phases. The direct evaluation 
of the path integral of a valence gluon confined by the stochastic background fields can be done by using 
parametrization~(\ref{smin}) for the minimal area. This parametrization reduces the path integral to that of the three-dimensional harmonic oscillator. In the deconfinement phase, at $T_c<T<T_{\rm d.r.}\simeq2T_c$, freezing 
of $\alpha_s(p)$ is present at Minkowskian $p^2<0$ due to the so-called magnetic confinement.
Upon the calculation of the path integral, we find the momentum scales at which the freezing occurs in the confinement and 
deconfinement phases. Analytically, these scales are given by the same formula~(\ref{ma}), 
where $d=4$ at $T=0$ and $d=3$ at $T>T_c$ (cf. Eq.~(\ref{MT})). Numerically, the values of the freezing scales following from 
the fits in the momentum representation are smaller than the corresponding analytic ones and closer to the phenomenological value of 1~GeV (see Eq.~(\ref{mass1}) and the end of Section~5). The values of $\alpha_s(0)$ obtained 
at $T=0$, $T_c$, $1.5T_c$, and $2T_c$ are 0.341, 0.377, 0.357, and 0.333, respectively. Full plots, which include the experimentally measured $\alpha_s(p)$, are presented in Fig.~5. Finally, we have estimated physical effect of the freezing on the thrust variable. The calculated nonperturbative contribution, which arises due to the soft radiation, brings 
the purely perturbative value of this quantity closer to the experimental one. 

Note in conclusion that one more application of the proposed method can be a calculation of mean sizes of the gluon bound states. Such bound states can be not only color-singlet (glueball), but also color-octet (quark-gluon). The latter
are suggested to be important ingredients of the quark-gluon plasma at temperatures $T_c<T<T_{\rm d.r.}$~\cite{sz, bs}.
Work in this direction is now in progress.

\section*{Acknowledgments}
\noindent
We are grateful to J.~Braun, A.~Di~Giacomo, H.~Gies, and Yu.A.~Simonov for useful discussions.
The work of D.A. has been supported through the contract MEIF-CT-2005-024196.

\section*{Appendix A. Details of calculation of $\Pi(x,y)$ at $T=0$.}
The path integrals in Eq.~(\ref{PI}) read as follows
$$
\int({\cal D}u_\mu)_{xy}\exp\left(-\frac{\mu+\bar\mu}{2}\int_0^R d\tau\dot u_\mu^2\right)=
\left(\frac{\mu+\bar\mu}{2\pi R}\right)^2\exp\left[-\frac{(\mu+\bar\mu)R}{2}\right],\eqno(A.1)
$$
$$\int({\cal D}r_4)_{00}
\exp\left(-\frac{\mu_r}{2}\int_0^R d\tau\dot r_4^2\right)=\sqrt{\frac{\mu_r}{2\pi R}},\eqno(A.2)$$
$$
\int({\cal D}{\bf r})_{00}
\exp\left(-\frac{\mu_r}{2}\int_0^R d\tau\dot {\bf r}^2-\frac{\sigma^2R}{4\lambda}\int_0^R 
d\tau{\bf r}^2\right)=
\left[\frac{\omega}{4\pi\sinh\left(\frac{\omega R}{2\mu_r}\right)}\right]^{3/2},\eqno(A.3)$$
where $\omega\equiv\sigma\sqrt{2\mu_r R/\lambda}$ is the frequency of the harmonic oscillator.
Bringing these formulae together and passing from the integration over $\lambda$ to the integration over
$\xi\equiv\sigma R^{3/2}/\sqrt{2\mu_r\lambda}$, we arrive at Eq.~(\ref{eqf}) as it would look like before the introduction of the variables $a$ and $b$:
$$
\Pi(x,y)\simeq\frac{1}{2^{11/2}\pi^{9/2}}\frac{\sigma}{\sqrt{R}}\int_0^\infty\frac{d\xi}{\sqrt{\xi}\sinh^{3/2}\xi}
\int_0^\infty\frac{d\mu d\bar\mu}{\sqrt{\mu_r}}
\exp\left[-\frac{\mu+\bar\mu}{2}R-\frac{\sigma^2R^3}{2\xi^2}
\left(\frac{1}{\mu}+\frac{1}{\bar\mu}\right)\right]. \eqno(A.4)
$$
It further turns out that the saddle-point integral over $\mu$ and $\bar\mu$ can be done
even with the account for $1/\sqrt{R\mu_r}$ in the pre-exponent. Indeed, we are dealing with the integral
$\int_0^\infty d\mu d\bar\mu{\rm e}^{-f(\mu,\bar\mu)}$, where 
$$
f(\mu,\bar\mu)\equiv\frac{\mu+\bar\mu}{2}R+\frac{\sigma^2R^3}{2\xi^2}
\left(\frac{1}{\mu}+\frac{1}{\bar\mu}\right)-\frac12\ln
\left[\frac{1}{R}\left(\frac{1}{\mu}+\frac{1}{\bar\mu}\right)\right].$$
The saddle-point equation $\frac{\partial f}{\partial\bar\mu}=0$ reads
$R\bar\mu^2+\mu_r-\frac{\sigma^2R^3}{\xi^2}=0$.
We can further use the fact that, since $f$ is symmetric under $\mu\leftrightarrow\bar\mu$, the saddle-point values of $\mu$ and $\bar\mu$ coincide. Therefore, setting in the last equation $\bar\mu=\mu$, we arrive at a quadratic
equation, whose solution is
$$
\mu=\bar\mu=\frac{1}{4R}\left[-1+\sqrt{1+\left(\frac{4\sigma R^2}{\xi}\right)^2}\right]. \eqno(A.5)
$$
The sign "$+$" in front of the square root has been chosen because the integration region over $\mu$ and $\bar\mu$ 
is from $0$ to $+\infty$. 

Further, as we are interested in the large-distance regime, $\sigma R^2\gtrsim 1$, we can consider separately
the following three regions of integration over $\xi$:

\noindent
$\bullet$
$\xi<1<4\sigma R^2$. At these values of $\xi$, the saddle-point values~(A.5) are $\mu=\bar\mu
\simeq\frac{\sigma R}{\xi}$. Accordingly, ${\rm e}^{-f}\Bigr|_{\rm saddle-point}=\frac{1}{R}\sqrt{\frac{2\xi}{\sigma}}{\rm e}^{-2\sigma R^2/\xi}$. Next, because $f$ is a symmetric 
function of $\mu$ and $\bar\mu$, the pre-exponent of the saddle-point integral reads 
$$\frac{2\pi}{\frac{\partial^2f}{\partial \bar\mu^2}\Bigr|_{\rm saddle-point}}\simeq\frac{2\pi\sigma}{\xi}\left(1+\frac38\frac{\xi}{\sigma R^2}\right).$$ 
Altogether, 
$$
\int_0^\infty d\mu d\bar\mu{\rm e}^{-f(\mu,\bar\mu)}\simeq\frac{2\pi}{R}
\sqrt{\frac{2\sigma}{\xi}}\left(1+\frac38\frac{\xi}{\sigma R^2}\right)
{\rm e}^{-2\sigma R^2/\xi}. \eqno (A.6)
$$
Further, approximating $\frac{1}{\sqrt{\xi}\sinh^{3/2}\xi}$ in Eq.~(A.4) by $\frac{1}{\xi^2}$ 
and changing the integration variable to $t=\frac{2\sigma R^2}{\xi}$,  
we have for the contribution to $\Pi(x,y)$ from this region of $\xi$'s
$$
\Pi^{(1)}(x,y)\simeq\frac{\sigma^{3/2}}{16\pi^{7/2}R}\left[\frac{1}{(2\sigma)^{3/2}R^3}\Gamma\left(\frac32,2\sigma R^2\right)+
\frac{3}{2^{7/2}\sigma^{3/2}R^3}\Gamma\left(\frac12,2\sigma R^2\right)\right],$$
where $\Gamma(\alpha,x)=\int_{x}^{\infty}dtt^{\alpha-1}{\rm e}^{-t}$ is the incomplete Gamma-function.
Using the known asymptotics $\Gamma(\alpha,x)\to x^{\alpha-1}{\rm e}^{-x}$ at $x>1$, we finally obtain 
$$
\Pi^{(1)}(x,y)\simeq
\frac{\sigma^{1/2}{\rm e}^{-2\sigma R^2}}{32\pi^{7/2}R^3}
\left(1+\frac{3}{8\sigma R^2}\right). \eqno (A.7)
$$

\noindent
$\bullet$
$1<\xi<4\sigma R^2$. In this region of $\xi$'s, Eq.~(A.6) still holds, but $\frac{1}{\sqrt{\xi}\sinh^{3/2}\xi}$ in Eq.~(A.4) should be approximated by 
$\frac{2^{3/2}}{\sqrt{\xi}{\rm e}^{3\xi/2}}$. This yields
$$
\Pi^{(2)}(x,y)\simeq\frac{\sigma^{3/2}(I_1+I_2)}{(32\pi^7)^{1/2}R},~ {\rm where}~  
I_1\equiv\int_{1}^{4\sigma R^2}\frac{d\xi}{\xi}{\rm e}^{-\frac{3\xi}{2}-\frac{2\sigma R^2}{\xi}},~
I_2\equiv\frac{3}{8\sigma R^2} \int_{1}^{4\sigma R^2}d\xi{\rm e}^{-\frac{3\xi}{2}-\frac{2\sigma R^2}{\xi}}. \eqno (A.8)
$$
Let us start with the analysis of $I_1$ by representing the integration region as
$\int_{1}^{4\sigma R^2}=\int_0^\infty-\int_0^1-\int_{4\sigma R^2}^\infty$. Then, the first of these three integrals 
can be done exactly and reads $2K_0(2\sqrt{3\sigma}R)$, where here and below $K_\nu$'s are the Macdonald functions.
We can further take into account that the saddle point of ${\rm e}^{-\frac{3\xi}{2}-\frac{2\sigma R^2}{\xi}}$,
which is $\xi=2\sqrt{\sigma/3}R$, lies between $1$ and $4\sigma R^2$. Owing to this fact, we can disregard $\frac{3\xi}{2}$ in the exponent of the integral $\int_0^1$, in the same way as we can disregard $\frac{2\sigma R^2}{\xi}$ in the exponent of the 
integral $\int_{4\sigma R^2}^\infty$. These approximations yield at $\sigma R^2>1$
$$
I_1\simeq2K_0(2\sqrt{3\sigma}R)-\Gamma(0,2\sigma R^2)-\Gamma(0,6\sigma R^2)\to\sqrt{\frac{\pi}{\sqrt{3\sigma}R}}{\rm e}^{-2\sqrt{3\sigma}R}-
\frac{{\rm e}^{-2\sigma R^2}}{2\sigma R^2}-\frac{{\rm e}^{-6\sigma R^2}}{6\sigma R^2}.$$
Analogously,
$$
I_2\simeq\frac{3}{8\sigma R^2}\left[4R\sqrt{\frac{\sigma}{3}}K_1(2\sqrt{3\sigma}R)-\int_0^1d\xi{\rm e}^{-2\sigma R^2/\xi}-
\int_{4\sigma R^2}^{\infty}d\xi{\rm e}^{-3\xi/2}\right]\to$$
$$\to\frac{3}{8\sigma R^2}\left[2\sqrt{\frac{\pi R}{3}
\sqrt{\frac{\sigma}{3}}}{\rm e}^{-2\sqrt{3\sigma}R}-\frac{{\rm e}^{-2\sigma R^2}}{2\sigma R^2}-\frac23\frac{{\rm e}^{-6\sigma R^2}}{\sigma R^2}\right]~ {\rm at}~ \sigma R^2>1.$$
Equation~(A.8) then yields:
$$
\Pi^{(2)}(x,y)\simeq\frac{\sigma^{3/2}}{(32\pi^7)^{1/2}R}\Biggl[\frac{\pi^{1/2}}{(3\sigma)^{1/4}R^{1/2}}\left(1+
\frac{1}{4R}\sqrt{\frac{3}{\sigma}}\right){\rm e}^{-2\sqrt{3\sigma}R}-$$
$$-\frac{1}{2\sigma R^2}\left(1+\frac{3}{8\sigma R^2}\right){\rm e}^{-2\sigma R^2}-
\frac{5}{12}\frac{{\rm e}^{-6\sigma R^2}}{\sigma R^2}\Biggr]. \eqno (A.9)
$$
\noindent
$\bullet$
$1<4\sigma R^2<\xi$. At these values of $\xi$, the saddle-point values of $\mu$ and $\bar\mu$, Eq.~(A.5), can be approximated as $\mu=\bar\mu\simeq\frac{2\sigma^2R^3}{\xi^2}$. The exponent 
at the saddle point reads ${\rm e}^{-f}\Bigr|_{\rm saddle-point}=\frac{\xi}{\sigma R^2}{\rm e}^{-\frac12-\frac{2(\sigma R^2)^2}{\xi^2}}$. Again, owing to the symmetry of $f$ under $\mu\leftrightarrow\bar\mu$,
we find $\frac{\partial^2f}{\partial \bar\mu^2}\Bigr|_{\rm saddle-point}\simeq\frac{\xi^4}{32\sigma^4R^6}$.  Therefore, the saddle-point integral over $\mu$ and $\bar\mu$ reads 
$$\int_0^\infty d\mu d\bar\mu{\rm e}^{-f(\mu,\bar\mu)}\simeq\frac{64\pi}{\sqrt{{\rm e}}}\frac{\sigma^3R^4}{\xi^3}{\rm e}^{-\frac{2(\sigma R^2)^2}{\xi^2}}.$$
Equation~(A.4) then yields for the contribution to the polarization operator, which comes about from this region of $\xi$'s: 
$$
\Pi^{(3)}(x,y)\simeq\frac{4(\sigma R)^4}{\sqrt{\pi^7{\rm e}}}\int_{4\sigma R^2}^\infty\frac{d\xi}{\xi^{7/2}}
{\rm e}^{-\frac{3\xi}{2}-\frac{2(\sigma R^2)^2}{\xi^2}}.$$
The value of the saddle point of the exponent, $\xi=\frac{2(\sigma R^2)^{2/3}}{3^{1/3}}$, is smaller than $4\sigma R^2$, 
for which reason we approximately have  
$$
\Pi^{(3)}(x,y)\simeq\frac{4(\sigma R)^4}{\sqrt{\pi^7{\rm e}}}\int_{4\sigma R^2}^\infty\frac{d\xi}{\xi^{7/2}}
{\rm e}^{-\frac{3\xi}{2}}=\sqrt{\frac{3^5}{2\pi^7{\rm e}}}(\sigma R)^4\Gamma\left(-\frac52,6\sigma R^2\right)
\simeq\frac{\sqrt{\sigma}}{48\sqrt{\pi^7{\rm e}}R^3}{\rm e}^{-6\sigma R^2}. \eqno (A.10)
$$
Bringing together Eqs.~(A.7), (A.9), and (A.10), we can write the final result for the 
polarization operator $\Pi(x,y)=\Pi^{(1)}(x,y)+\Pi^{(2)}(x,y)+\Pi^{(3)}(x,y)$:
$$
\Pi(x,y)\simeq\frac{\sigma^{5/4}}{2^{5/2}\cdot3^{1/4}\cdot\pi^3\cdot R^{3/2}}
\left(1+\frac{1}{4R}\sqrt{\frac{3}{\sigma}}\right){\rm e}^{-2\sqrt{3\sigma}R}+{\cal O}\left({\rm e}^{-2\sigma R^2}\right)+
{\cal O}\left({\rm e}^{-6\sigma R^2}\right),$$
where  
$$
{\cal O}\left({\rm e}^{-2\sigma R^2}\right)\equiv\frac{1-2^{3/2}}{32\pi^{7/2}}\frac{\sigma^{1/2}}{R^3}
\left(1+\frac{3}{8\sigma R^2}
\right){\rm e}^{-2\sigma R^2}, {\cal O}\left({\rm e}^{-6\sigma R^2}\right)\equiv
\frac{{\rm e}^{-1/2}-5\cdot 2^{-1/2}}{48\pi^{7/2}}
\frac{\sigma^{1/2}}{R^3}{\rm e}^{-6\sigma R^2}.$$
Note that the correction ${\cal O}\left({\rm e}^{-6\sigma R^2}\right)$ is negative.
The leading term, ${\cal O}\left({\rm e}^{-2\sqrt{3\sigma}R}\right)$, comes about from Eq.~(A.9).

\section*{Appendix B. Details of calculation of $\Pi(x,y)$ at $T>T_c$.}
Let us start with the polarization operator at $T>T_{\rm d.r.}$. 
Choosing for concreteness 
${\bf R}=(L,0,0)$, we define the relative coordinate ${\bf r}(\tau)\equiv{\bf z}(\tau)-\bar{\bf z}(\tau)=(r_1(\tau),\vec\rho{\,}(\tau))$ and the "center-of-mass" coordinate 
${\bf u}=\frac{\bar s{\bf z}+s\bar{\bf z}}{s+\bar s}$. After changing the variables $(s, \bar s)$ to $(\mu, \bar\mu)$
according to Eq.~(\ref{mubarmu}), we have for polarization operator~(\ref{PiT}) [cf. Eq.~(\ref{PI})]:
$$
\Pi({\bf x},{\bf y}|T)\simeq\left(\frac{LT}{2}\right)^2\int_0^\infty\frac{d\mu}{\mu^2}
\int_0^\infty\frac{d\bar\mu}{\bar\mu^2}I(\mu,\bar\mu),~~ {\rm where} \eqno(B.1)$$

$$I(\mu,\bar\mu)=\int({\cal D}{\bf u})_{{\bf x}{\bf y}}
({\cal D}{\bf r})_{00}
\int_0^\infty\frac{d\lambda}{\sqrt{\pi\lambda}}
\exp\Biggl[-\lambda
-\frac{\mu+\bar\mu}{2}\int_0^Ld\tau\dot{\bf u}^2-\frac{\mu_r}{2}\int_0^Ld\tau\dot{\bf r}^2-
\frac{\sigma_s^2L}{4\lambda}\int_0^Ld\tau\vec\rho{\,}^2\Biggr].$$
Carrying out the path integrations over ${\bf r}$ and ${\bf u}$, we have
$$
I(\mu,\bar\mu)=\frac{\sigma_s\mu\bar\mu\sqrt{\mu+\bar\mu}}{8\sqrt{2}\pi^{7/2}L^{3/2}}
\exp\left[-\frac{(\mu+\bar\mu)L}{2}\right]\int_0^\infty\frac{d\lambda}{\lambda}
\frac{{\rm e}^{-\lambda}}{\sinh\left(\sigma_s\sqrt{\frac{L^3}{2\mu_r\lambda}}\right)}. \eqno(B.2)
$$
Changing the integration variable $\lambda\to\xi=\sigma_s\sqrt{\frac{L^3}{2\mu_r\lambda}}$, one can rewrite this equation as
$$
I(\mu,\bar\mu)=\frac{\sigma_s}{4\sqrt{2\pi^7L^3}}\mu\bar\mu\sqrt{\mu+\bar\mu}\int_{0}^{\infty}
\frac{d\xi}{\xi\sinh\xi}\exp\left\{-\frac{L}{2}\left[\mu+\bar\mu+\left(\frac{\sigma_sL}{\xi}\right)^2\left(
\frac{1}{\mu}+\frac{1}{\bar\mu}\right)\right]\right\}. \eqno(B.3)
$$
One should now substitute this expression into Eq.~(B.1) and perform approximate 
saddle-point integrations over $\mu$ and $\bar\mu$. This procedure means the insertion of the 
saddle-point values into the pre-exponent. In this way, one arrives at the following counterpart of Eq.~(\ref{Papr}) at 
$T>T_{\rm d.r.}$:
$$
\Pi({\bf x},{\bf y}|T)\simeq\frac{\sqrt{\sigma_s}T^2}{8\pi^{5/2}L}\int_0^\infty\frac{d\xi}{\sqrt{\xi}\sinh\xi}
{\rm e}^{-2\sigma_sL^2/\xi}. \eqno(B.4)
$$
Note that, in three dimensions, $\sinh^{\frac{d-1}{2}}\xi=\sinh\xi$ as explained after Eq.~(\ref{ma}).
By splitting the integration region as in Eq.~(\ref{split}), the last integral can be approximated as $I_1+I_2$, where 
$$I_1\equiv\int_0^1\frac{d\xi}{\xi^{3/2}}{\rm e}^{-2\sigma_s L^2/\xi},~~ 
I_2\equiv2\int_1^\infty\frac{d\xi}{\sqrt{\xi}}{\rm e}^{-\xi-2\sigma_s L^2/\xi}.$$
To evaluate these integrals notice that, at temperatures $T>T_{\rm d.r.}$, which we are currently considering,  
the spatial string tension is parametrically $\sigma_s\propto g_3^2\sim T^2$. Numerically, the ratio 
$\sigma_s/\sigma_0$ is larger than 2 at $T>T_{\rm d.r.}$ (cf. Fig.~4). 
Next, for magnetic confinement to hold, the spatial Wilson loop should be sufficiently large, in particular the distance $L$ should be $\gtrsim 1{\,}{\rm fm}$.  For such distances, $2\sigma_s L^2>4\sigma_0L^2\gg1$, and we have 
$$I_1\simeq\frac{{\rm e}^{-2\sigma_s L^2}}{2\sigma_s L^2},~~   
I_2\simeq2\sqrt{\pi}{\rm e}^{-2\sqrt{2\sigma_s}L}-\frac{{\rm e}^{-2\sigma_s L^2}}{\sigma_s L^2}.$$
Bringing these expressions together and neglecting the subleading terms $\sim\frac{{\rm e}^{-(mL/2)^2}}{(mL)^2}$, 
we arrive at Eq.~(\ref{polTfinal}).

In the opposite limit of small distances, Eq.~(B.2) yields 
$$
I(\mu,\bar\mu)\to\frac{(\mu\bar\mu)^{3/2}}{(2\pi L)^3}\exp\left[-\frac{(\mu+\bar\mu)L}{2}\right]. \eqno(B.5)
$$
Inserting this expression into Eq.~(B.1), we have
$$
\Pi({\bf x},{\bf y}|T)\to\frac{T^2}{32\pi^3L}\int_0^\infty\frac{d\mu}{\sqrt{\mu}}\int_0^\infty\frac{d\bar\mu}{\sqrt{\bar\mu}}
\exp\left[-\frac{(\mu+\bar\mu)L}{2}\right]~~ {\rm at}~ \sigma_sL^2\ll 1.
$$
Straightforward integrations over $\mu$ and $\bar\mu$ in this formula lead to Eq.~(\ref{PhighT}).

Let us now consider the temperature interval $T_c<T<T_{\rm d.r.}$. Changing in Eq.~(\ref{genE}) the variables $(s, \bar s)$ 
to $(\mu, \bar\mu)$, we have
$$
\Pi({\bf x},{\bf y}|T)=\frac{L}{8\pi}\int_0^\infty\frac{d\mu}{\mu^{3/2}}\int_0^\infty\frac{d\bar\mu}{\bar\mu^{3/2}}I(\mu,\bar\mu)
\sum\limits_{n,k}^{}\exp\left\{-\frac{1}{2L}\left[\mu(R_4-\beta n)^2+\bar\mu(R_4-\beta k)^2\right]\right\}. \eqno(B.6)
$$
Using for $I(\mu,\bar\mu)$ representation~(B.3) and performing the saddle-point integrations over $\mu$, $\bar\mu$ as above,
we arrive at the following analogue of Eq.~(B.4):
$$\Pi(x,y|T)\simeq\frac{\sigma_s^{3/2}}{16\pi^3\sqrt{2\pi}L}\int_0^\infty\frac{d\xi}{\xi^{3/2}\sinh\xi}\sum\limits_{n,k}^{}
\sqrt{\frac{\varphi_n+\varphi_k}{(\varphi_n\varphi_k)^3}}{\rm e}^{-\frac{\sigma_sL^2}{\xi}(\varphi_n+\varphi_k)},~~ {\rm where}$$
$$\varphi_n\equiv\sqrt{1+\left(\frac{R_4-\beta n}{L}\right)^2}.$$
When splitting the $\xi$-integral as in Eq.~(\ref{split}),
$$
\int_0^\infty\frac{d\xi}{\xi^{3/2}\sinh\xi}{\rm e}^{-\frac{\sigma_sL^2}{\xi}(\varphi_n+\varphi_k)}\simeq
\int_{0}^{1}\frac{d\xi}{\xi^{5/2}}
{\rm e}^{-\frac{\sigma_sL^2}{\xi}(\varphi_n+\varphi_k)}+2\int_{1}^{\infty}\frac{d\xi}{\xi^{3/2}}{\rm e}^{-\xi
-\frac{\sigma_sL^2}{\xi}(\varphi_n+\varphi_k)}, \eqno(B.7)
$$
we notice that, at temperatures $T_c<T<T_{\rm d.r.}$ of interest, the ratio $\sigma_s/\sigma_0$ ranges 
between 1 and 2 (see Fig.~4). Therefore, 
$$\sigma_sL^2(\varphi_n+\varphi_k)\ge2\sigma_sL^2>2\sigma_0L^2\gg 1~~ {\rm at}~~ L\gtrsim1{\,}{\rm fm}.$$ 
For this reason,
the second of the two integrals on the R.H.S. of Eq.~(B.7) is again saturated by its saddle point, 
$\xi_{\rm s.p.}=\sqrt{\sigma_sL^2(\varphi_n+\varphi_k)}$, and 
yields the leading exponential fall-off with $L$, whereas the first integral yields a subleading Gaussian term.
Disregarding the latter, we have for the leading exponential fall-off:
$$\Pi(x,y|T)\simeq\frac{\sigma_s}{8\sqrt{2}\pi^3L^2}
\sum\limits_{n,k}^{}\frac{{\rm e}^{-2\sqrt{\sigma_sL^2(\varphi_n+\varphi_k)}}}{(\varphi_n\varphi_k)^{3/2}}.$$
Now, as we have just seen, the argument of the exponent here is much larger than 1. Therefore, one can safely restrict oneself to the $\varphi_0$-terms in the sums over $n$ and $k$. Since $\varphi_0=R/L$, where $R\equiv\sqrt{R_4^2+L^2}$, we arrive at
Eq.~(\ref{ld}).

In the opposite case of small distances, inserting Eq.~(B.5) into Eq.~(B.6), we have
$$\Pi(x,y|T)\to\frac{1}{64\pi^4L^2}\sum\limits_{n,k}^{}\int_0^\infty d\mu\int_0^\infty d\bar\mu
\exp\left[-\frac{L}{2}\left(\mu\varphi_n^2+\bar\mu\varphi_k^2\right)\right]=$$
$$
=\frac{1}{16\pi^4}\sum\limits_{n,k}^{}\frac{1}{\left[L^2+(R_4-\beta n)^2\right]\left[L^2+(R_4-\beta k)^2\right]}~~ {\rm at}~ \sigma_sR^2\ll 1. \eqno(B.8)
$$
Therefore, we have recovered in the short-distance limit the product of two free scalar thermal propagators. 
Note that the sums in Eq.~(B.8) can be done analytically. Setting $L=R$, $R_4=0$, we obtain Eq.~(\ref{pOl}).

\end{document}